\documentclass[11pt]{iopart}
\usepackage{epsfig}

\def\piz{$\pi^{0}$ }
\def\et{$\eta$ }
\def\roots{$\sqrt{s_{\mathrm{NN}}}$}
\def\raa{$R_{AA}$ }
\def\pt{$p_{\mathrm{T}}$ }

\begin{document}

\title{Diagnosing Energy Loss: PHENIX Results on High-$p_{\mathrm{T}}$ Hadron Spectra}
\author{B. Sahlmueller for the PHENIX\footnote{For the full list of
    PHENIX authors and acknowledgements, see Appendix
    'Collaborations' of this volume} Collaboration}
\address{University of M\"unster, Institut f\"{u}r Kernphysik,
  Wilhelm-Klemm-Str. 9, 48149 M\"{u}nster, Germany}
\ead{sahlmul@nwz.uni-muenster.de}

\begin{abstract}Measurements of inclusive spectra of hadrons at large
  transverse momentum over a broad range of energy in different
  collision systems have been performed with the PHENIX experiment at
  RHIC. The data allow to study the energy and system size dependence
  of the suppression observed in \raa of high-\pt hadrons at \roots =
  200 GeV. Due to the large energy
  range from \roots = 22 GeV to 200 GeV, the results can be
  compared to results from CERN SPS at a similar energy. The large
  Au+Au dataset from
  the 2004 run of RHIC also allows to constrain theoretical models
  that describe the hot and dense matter produced in such
  collisions. Investigation of particle ratios such as \et/\piz
  helps understanding the mechanisms of energy loss.\end{abstract}

\section{Introduction}

Previous measurements at RHIC have shown a significant suppression at
high \pt of
$\pi^{0}$, \et, and charged hadrons in central Au+Au collisions at \roots 
= 200 GeV compared to binary scaled p+p collisions
\cite{pi0suppr,suppr_star}. This
suppression was found to be \pt-independent for \pt $>$ 5 GeV/$c$
\cite{maya}. A suppression of \raa has also been observed in the smaller Cu+Cu
collision system, being similar in amount for similar numbers of
participants \cite{maya}.\\
Possible initial state effects
have been studied by measuring \piz production in d+Au collisions and
investigating a centrality dependence in such
collisions. These effects have been found to be small \cite{ppg044}, thus
the suppression in Au+Au is attributed to final state interactions such as
gluon radiation of partons in hot dense matter.\\
To shed further light on the mechanism of energy loss, the energy
dependence of the suppression can be studied with a variety of data as
RHIC delivers collision energies over a broad
range from \roots = 22.4 GeV which is close to SPS energies, up to
\roots = 200 GeV. Improved analysis methods and better statistics also
lead to a better understanding of energy loss using high-\pt hadron
spectra.\\

\section{Spectra measurement and $R_{AA}$}

The PHENIX experiment measures \piz and \et mesons via
their two-photon-decay. The decay photons are measured with the
Electromagnetic Calorimeter, consisting of two sectors of lead glass and six
sectors of lead scintillator sandwich calorimeters, at midrapidity. Each sector covers 22.5
degrees in azimuth. Uncorrected particle yields are extracted with an
invariant mass analysis using event mixing for background subtraction.
They are then corrected for
different effects such as the detector acceptance and the
reconstruction efficiency \cite{pi0suppr, ppg044, ppg055}.\\
\begin{figure}[t]
\begin{minipage}[t]{55mm}
\includegraphics[width=2.4in]{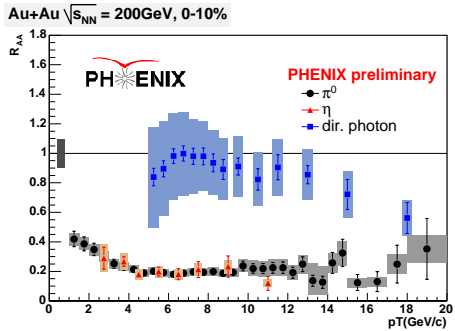}
\end{minipage}
\hspace{\fill}
\begin{minipage}[b]{100mm}
\caption{\label{fig:raa_auau}Nuclear modification factor \raa for \piz
  \cite{maya} and \et  mesons in comparison with direct photons
  \cite{tadaaki} in 0-10\% most central Au+Au collisions at \roots =
  200 GeV. The error bars show the \pt uncorrelated errors, the boxes
  around the points show the \pt correlated errors, the box at the
  left shows the normalization uncertainty.}
\end{minipage}
\end{figure}
The nuclear modification factor \raa 
%is a measure to compare the
%particle spectra in nucleus nucleus collisions with the ones in p+p
%collisions. It
is shown in Fig. \ref{fig:raa_auau} for \et mesons in Au+Au collisions
at \roots = 200 GeV in
comparison with \raa for \piz \cite{maya} and direct photons
\cite{tadaaki}. The suppression of both mesons appears to be the
same which can be explained with partonic energy loss in the
medium. The direct photons show an indication for suppression at \pt
$>$ 14 GeV/$c$. This is consistent with initial state effects
\cite{tadaaki}.\\
The energy dependence of \raa is examined in Cu+Cu collisions, where the
\piz \raa has been measured at three different energies (\roots =
200 GeV, 62.4 GeV, and 22.4 GeV), shown in Fig.
\ref{fig:raa_cucu}a for the most central events. The suppression is
strongest at high energies, \raa $\sim$ 0.4 at \roots = 200 GeV, while towards lower
energies, a Cronin like enhancement at \pt $<$ 5 GeV/$c$ becomes
clearly visible. Compared to Pb+Pb 
data from the WA98 experiment at CERN, measured at \roots = 17.2 GeV,
the 22.4 Cu+Cu \raa is similar for similar $N_{\mathrm{part}}$ as shown in Fig.
\ref{fig:raa_cucu}b. At this low energy, PHENIX has observed
no significant centrality dependence of the shape of the \piz spectra
(not shown).
\begin{figure}[t]
\begin{minipage}[b]{75mm}
\includegraphics[height=1.65in]{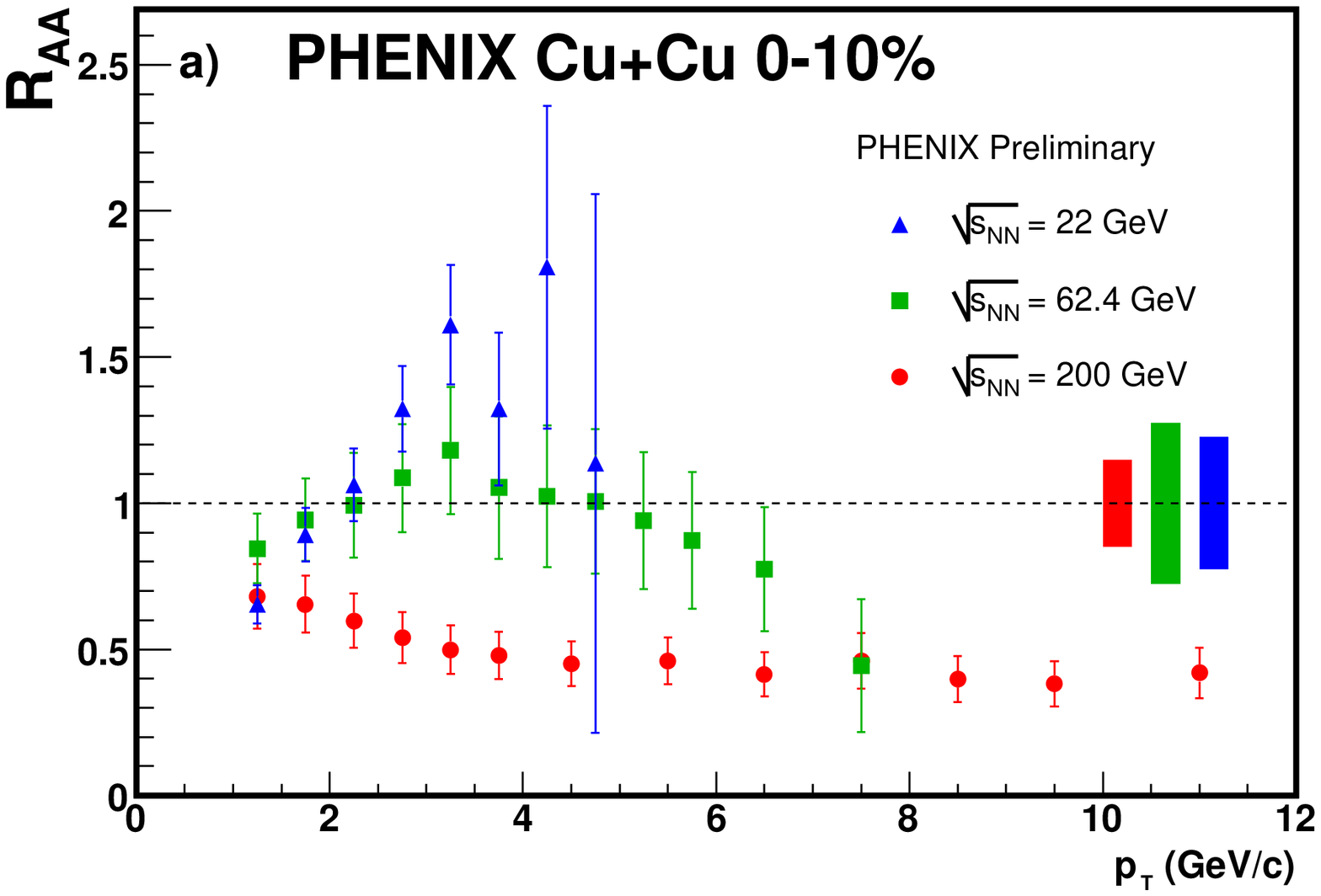}
\end{minipage}
\hspace{\fill}
\begin{minipage}[b]{75mm}
\includegraphics[height=1.65in]{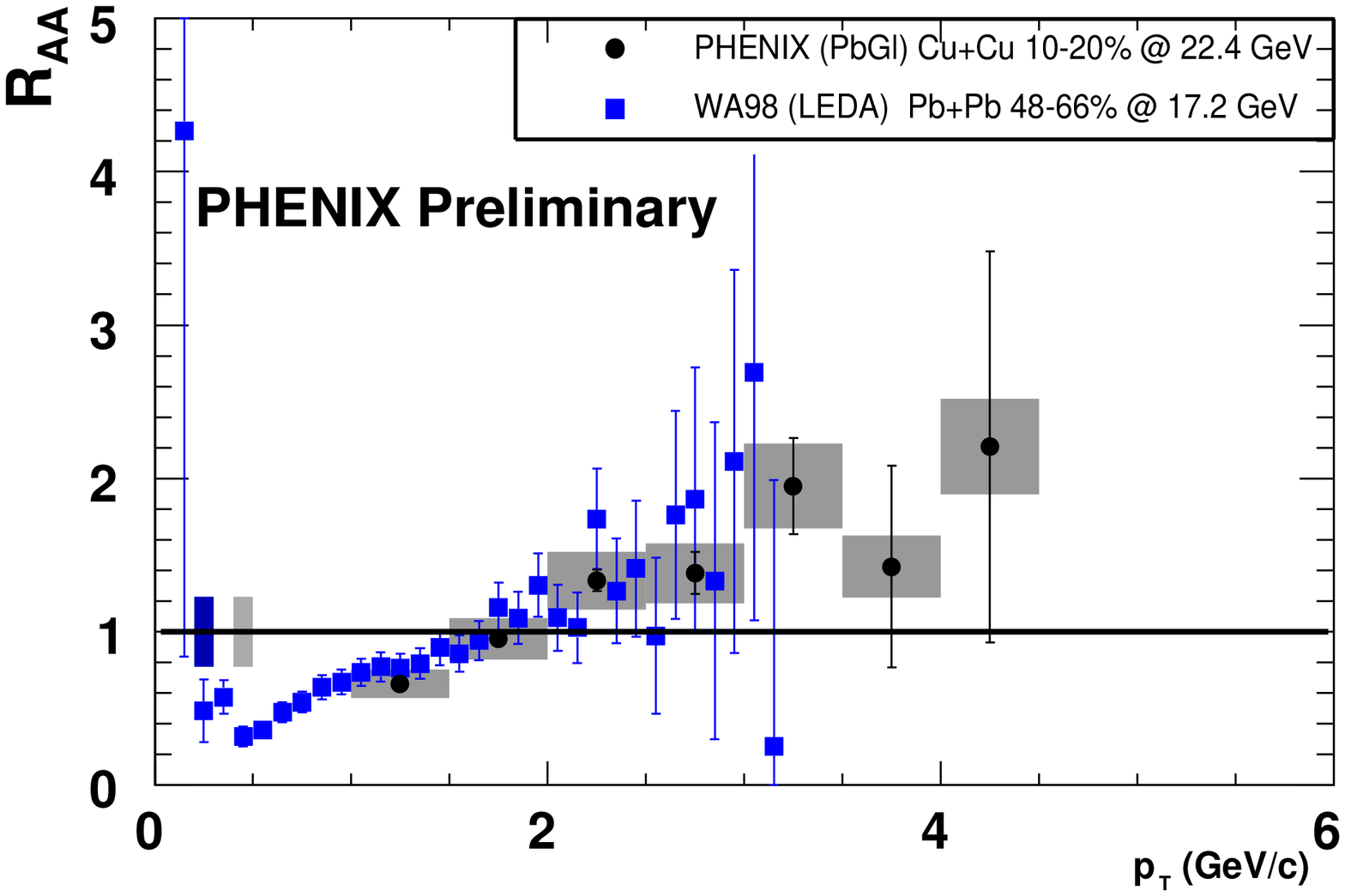}
\end{minipage}
\caption{\label{fig:raa_cucu}a) \piz \raa for
  the 0-10\% most central Cu+Cu collisions at \roots = 200 GeV,
  62.4 GeV, and 22.4 GeV. b) \piz \raa
  for 10-20 \% most central Cu+Cu
  ($N_{\mathrm{part}}=67.8$, \roots = 22.4 
  GeV) from PHENIX and for 48-66 \% most central Pb+Pb
  ($N_{\mathrm{part}}=63$, \roots = 17.2 GeV) from WA98
  \cite{WA98piz,WA98}. The error bars are of the same type as in
  Fig.~\ref{fig:raa_auau}.}
\end{figure}

\section{$\eta/\pi^{0}$ Ratio}
\begin{figure}[b]
\begin{minipage}[t]{75mm}
\includegraphics[height=1.65in]{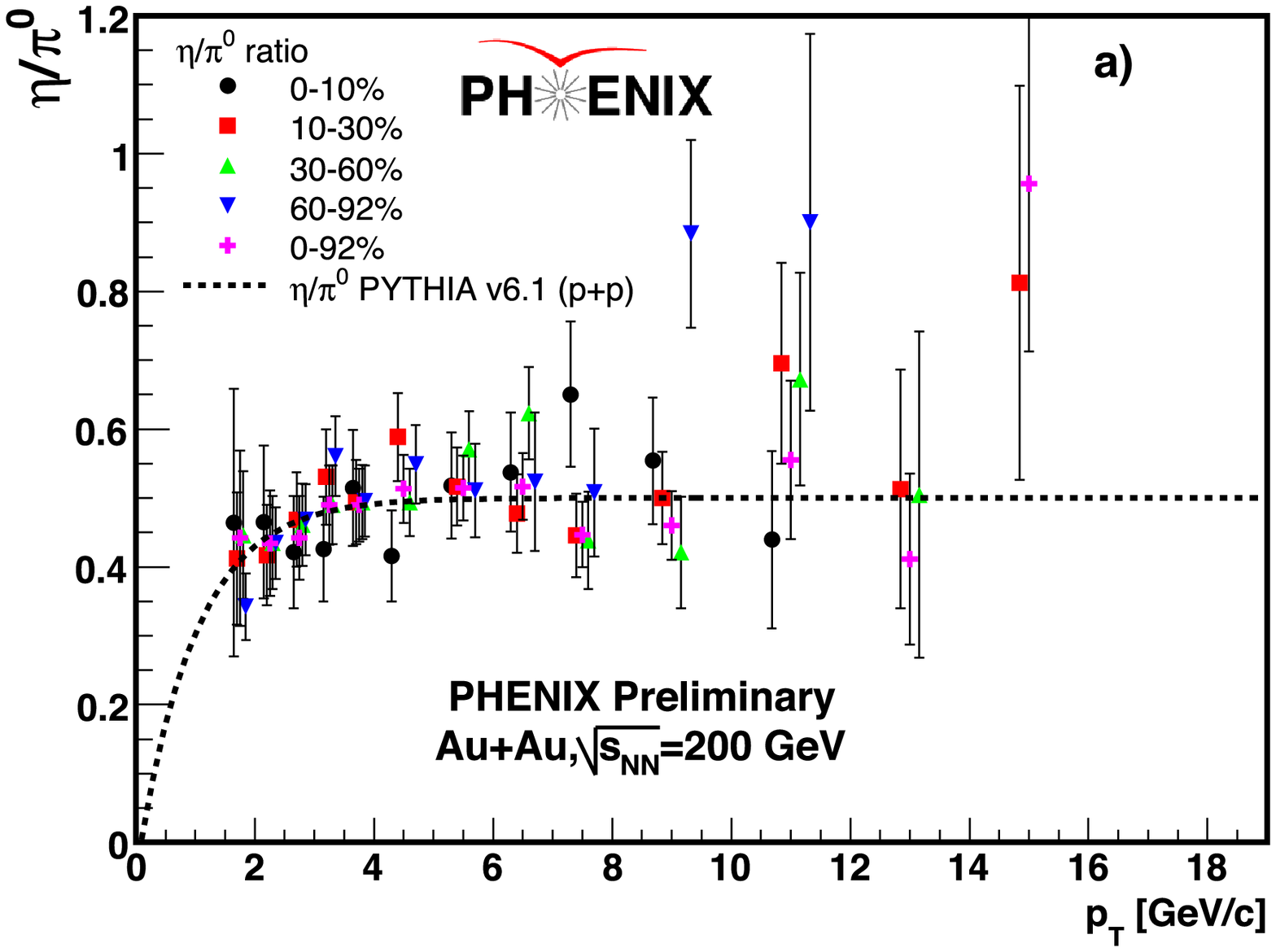}
\end{minipage}
\hspace{\fill}
\begin{minipage}[t]{75mm}
\includegraphics[height=1.65in]{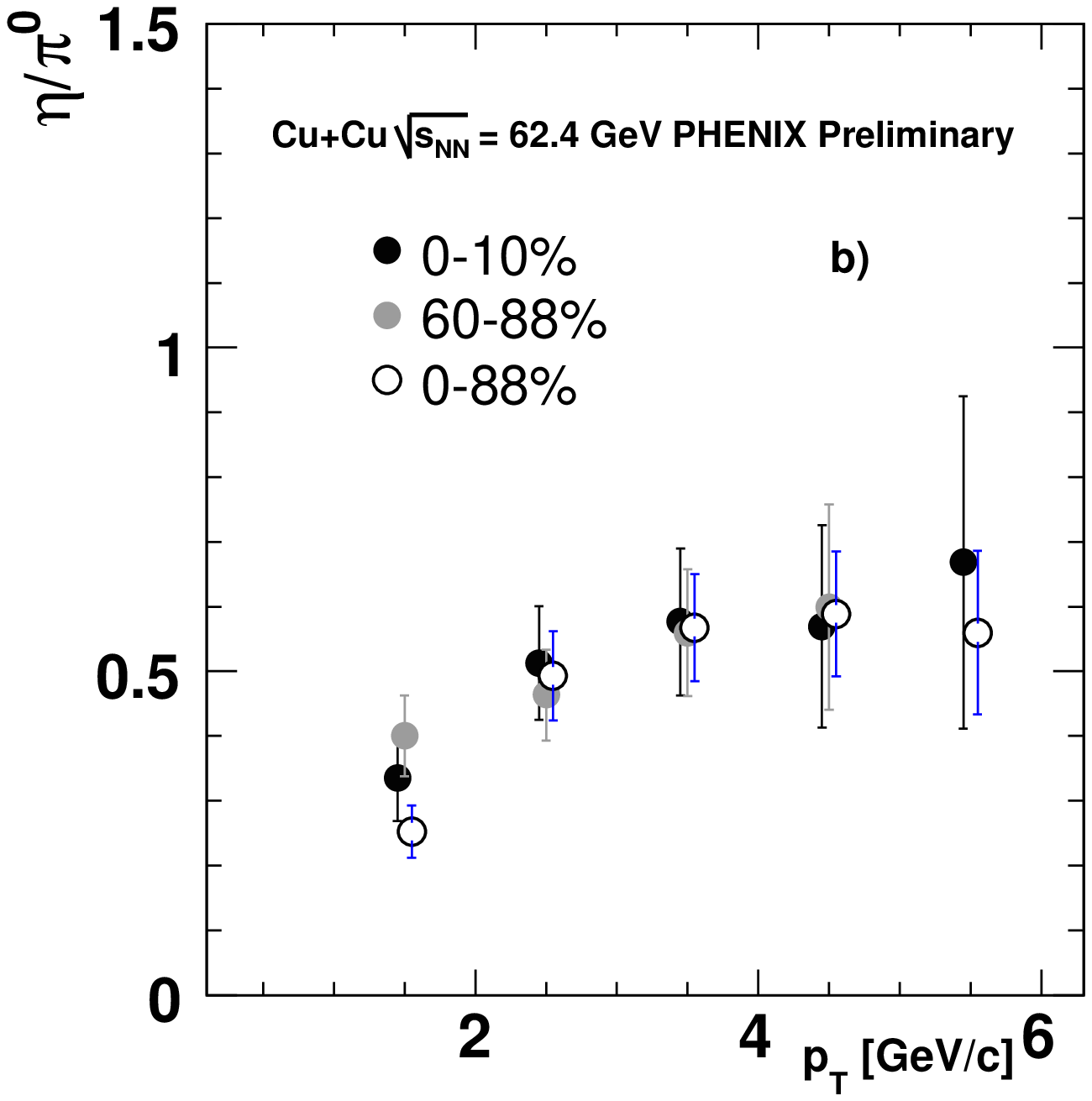}
\end{minipage}
\caption{\label{fig:etapi}Ratio of \et and \piz a) in Au+Au at
  \roots = 200 GeV for different centrality selections in comparison
  with a PYTHIA \cite{pythia} calculation and b) in Cu+Cu at \roots =
  62.4 GeV. The error bars show the total errors.}
\end{figure}

Fig. \ref{fig:etapi}a shows the \et/\piz ratio for 
Au+Au collisions at \roots = 200~GeV for different centrality
selections in comparison with a PYTHIA \cite{pythia}
calculation and in Fig.~\ref{fig:etapi}b 
for Cu+Cu collisions at \roots = 62.4 GeV. The ratio is found to
be independent of centrality over the whole \pt range and the PYTHIA
curve is in good agreement with the measured data. The ratio does
not show a system size or energy dependence and is consistent with
data from earlier measurements at different energies and collision
systems \cite{ppg055}. A possible explanation is that
the suppression of high-\pt hadrons occurs at the partonic level and 
that the fragmentation is not affected by the medium.

\section{Constraining Model Parameters}

\begin{figure}[t]
\begin{minipage}[b]{85mm}
\includegraphics[width=4in]{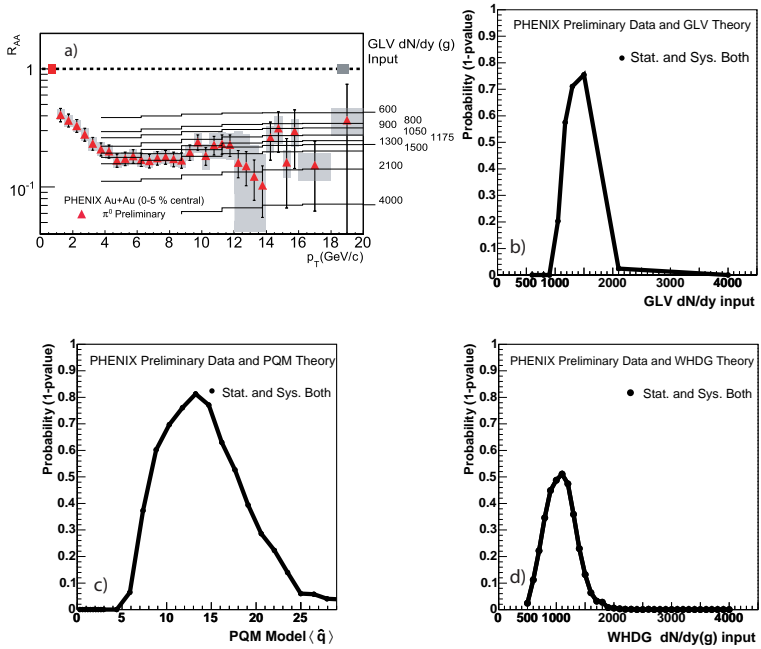}
\end{minipage}
\hspace{\fill}
\begin{minipage}[b]{70mm}
\caption{\label{fig:constr_theo}a) \piz \raa in Au+Au at \roots = 200
  GeV compared with predictions from a theoretical model \cite{glv}
  for different values of the initial gluon density $dN/dy$. b)
  $1-\mathrm{p}$-value for certain values of $dN/dy$ in GLV model. c)
  $1-\mathrm{p}$-value for certain values of medium transport
  coefficient $<$\^q$>$ in PQM model \cite{qhat}. d)
  $1-\mathrm{p}$-value for certain values of $dN/dy$ in WHDG model
  \cite{whdg}.}
\end{minipage}
\end{figure}
There are different theoretical models describing the observed
suppression of $R_{AA}$ at \roots = 200 GeV. These models describe the suppression as
function of the
initial gluon density $dN/dy$ \cite{glv,whdg} or the medium transport coefficient
$<$\^q$>$ \cite{qhat}. Both the
PHENIX measurement of the nuclear modification factor in Au+Au
collisions at \roots = 200 GeV and
theoretical predictions for different values of
the parameter $dN/dy$ are shown for one example \cite{glv}
in Fig. \ref{fig:constr_theo}a. The data can be used to constrain these
theoretical parameters. For this both the
correlated and the uncorrelated errors of the measurement are taken
into account in a $\chi^{2}$ analysis. For each value of the theory
parameter, in a first step the data 
points are varied within 4 RMS of the correlated errors to find the most
probable variation, taking the probability for both the offset and the
point-by-point deviation into account. In a second step, the
probabilities for numerous 
randomly chosen combinations of both error types are calculated. The
p-value is then defined as the fraction of these variations that are
less probable 
than the variation found in step one, so p-value $<$ 1. The
probabilities for different 
parameter choices are eventually calculated 
as $1-\mathrm{p}$-value, they are shown in Fig. \ref{fig:constr_theo}b, c, and d.

\section{Summary}
The PHENIX experiment has measured \piz and \et mesons over a
broad range of energies in different collision systems. The nuclear
modification factor of both mesons, which are suppressed by the same
factor, can be described with models
explaining this suppression with partons losing energy in the hot and
dense medium. The suppression shows an energy dependence and
becomes strongest at higher collision energies.\\
The production ratio \et/\piz is the same for different collision
systems and energies. It is unaffected by medium effects in different
collision systems. This observation supports the assumption of partonic
energy loss and fragmentation outside the medium.\\
The experimental data have been used to constrain the parameters of
different theoretical models. This is a first step to use the data for
more quantitative statements. However, the large uncertainties limit the 
discriminative power of the comparison.

\end{document}